\documentclass[runningheads]{llncs}
\usepackage{graphicx}
\usepackage[numbers,sort&compress]{natbib}

\begin{document}
\title{Extraction of Pulmonary Airway in CT Scans Using Deep Fully Convolutional Networks}
%
%
\author{Shaofeng Yuan}
\authorrunning{F. Author et al.}
%
\institute{School of Biomedical Engineering, Southern Medical University, Guangzhou, China \email{shaofeng.yuan.smu@gmail.com}}
\maketitle              
\begin{abstract}
Accurate, automatic and complete extraction of pulmonary airway in medical images plays an important role in analyzing thoracic CT volumes such as lung cancer detection, chronic obstructive pulmonary disease (COPD), and bronchoscopic-assisted surgery navigation. However, this task remains challenges, due to the complex tree-like structure of the airways. In this technical report, we use two-stage fully convolutional networks (FCNs) to automatically segment pulmonary airway in thoracic CT scans from multi-sites. Specifically, we firstly adopt a 3D FCN with U-shape network architecture to segment pulmonary airway in a coarse resolution in order to accelerate medical image analysis pipeline. And then another one 3D FCN is trained to segment pulmonary airway in a fine resolution. In the 2022 MICCAI Multi-site Multi-domain Airway Tree Modeling (ATM) Challenge, the reported method was evaluated on the public training set of 300 cases and independent private validation set of 50 cases. The resulting Dice Similarity Coefficient (DSC) is 0.914 $\pm$ 0.040, False Negative Error (FNE) is 0.079 $\pm$ 0.042, and False Positive Error (FPE) is 0.090 $\pm$ 0.066 on independent private validation set.

\end{abstract}
\section{Introduction}
Recently, fully convolutional networks (FCNs) were increasingly used in medical image segmentation~\cite{ronneberger2015,milletari2016}, such as U-Net and V-Net. For extraction of pulmonary airway in CT scans, FCNs-based methods~\cite{charbonnier2017,meng2017,nadeem2019,qin2019,wang2019,yun2019,zhao2019,qin2020,selvan2020,garcia-uceda2021,nadeem2021,qin2021,tan2021,zhang2021ijcars,zhang2021miccai,zheng2021miccai,zheng2021tmi,guo2022,huang2022,wang2022,wu2022,yu2022} were proposed and proved superior to previous approaches in~\cite{lo2012}. The organizers of the 2022 MICCAI Multi-site Multi-domain Airway Tree Modeling (ATM) Challenge collected 500 CT scans from multi-sites, i.e., the public LIDC-IDRI dataset and the Shanghai Chest hospital. In ATM 2022$\footnote{https://atm22.grand-challenge.org}$, 300 CT volumes are used for training$\footnote{actually, 299 cases are valid, and case 164 is discarded.}$, 50 CT volumes are used for validation, and 150 private CT volumes are used for testing.
\section{Method}
\begin{figure} \centering
\includegraphics[width=\textwidth]{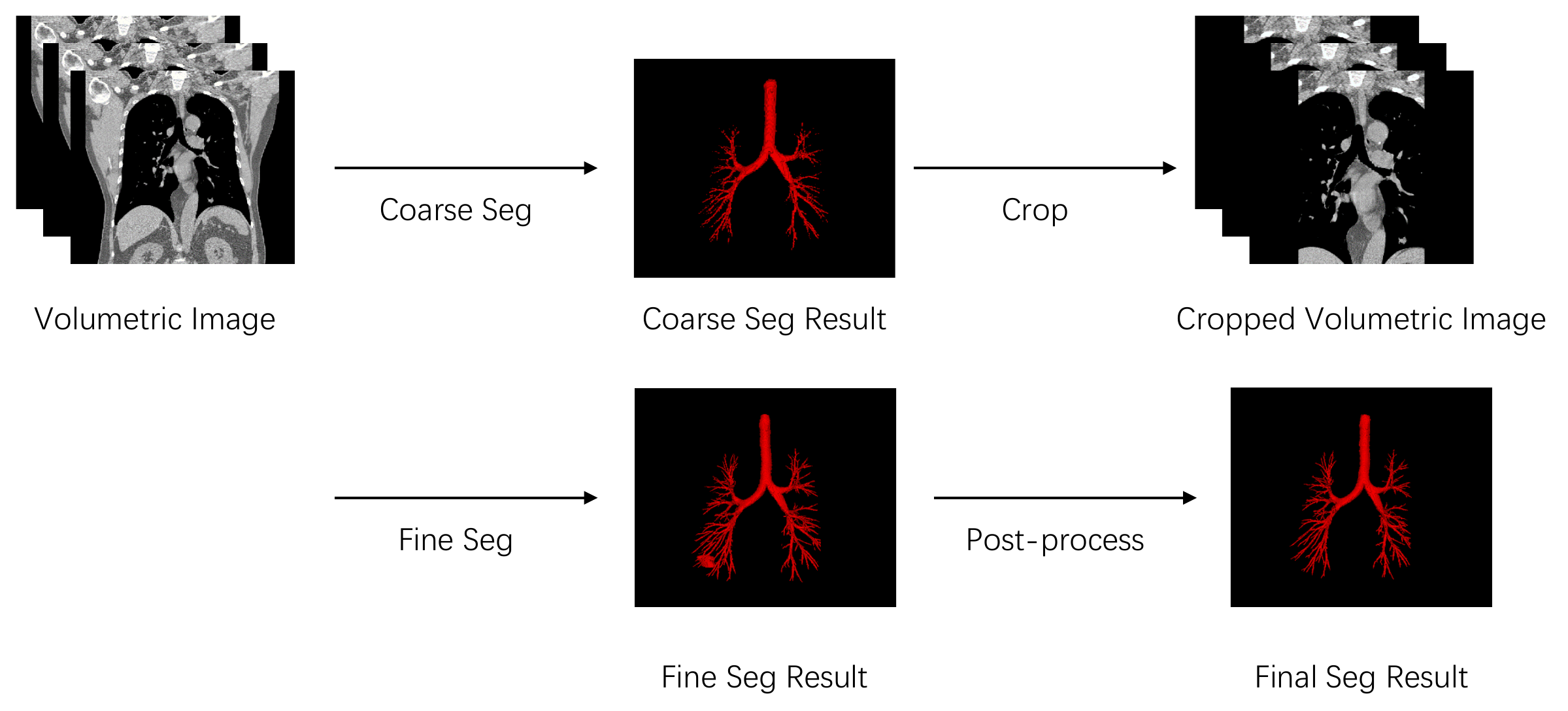}
\caption{The overview of the proposed \textbf{AirwaySeg} method.} \label{fig1}
\end{figure}

The overview of the proposed \textbf{AirwaySeg} method is shown in Fig.~\ref{fig1}.

\subsection{Coarse Segmentation}
Coarse segmentation of pulmonary airway is in order to accelerate medical image analysis pipeline. Thoracic CT volumes usually have very large image size. If 3D FCNs are used directly on original resolution, the processing time of one case is a few minutes. We adopt nnU-Net~\cite{isensee2021} as the coarse segmentation model.

\subsection{Fine Segmentation}
Fine segmentation of pulmonary airway in the extended bounding box from coarse segmentation is in order to extract airway tree as complete as possible. For convenience, we don't use 3D FCNs along the centerlines of pulmonary airway, but we use sliding window strategy. Also, We adopt nnU-Net~\cite{isensee2021} as the fine segmentation model.

\section{Result}
\subsubsection{Quantitative result in coarse segmentation stage} Table~\ref{tab1} gives a quantitative result of coarse segmentation model.

\begin{table} \centering
\caption{Quantitative results of \textbf{AirwaySeg}.}\label{tab1}
\begin{tabular}{|l|l|l|l|l|l|l|}
\hline
 & Dataset & Scans & Dice & Jaccard & Recall & Precision\\
\hline
Coarse Seg & validation & 60 & 0.832±0.041 & 0.715±0.059 & 0.771±0.062 & 0.908±0.033\\
Coarse Seg & train & 239 & 0.838±0.037 & 0.722±0.055 & 0.779±0.062 & 0.909±0.029\\
Fine Seg & validation & 60 & 0.912±0.064 & 0.844±0.092 & 0.949±0.025 & 0.885±0.098\\
Fine Seg & train & 239 & 0.911±0.045 & 0.840±0.071 & 0.947±0.031 & 0.882±0.075\\
\hline
\end{tabular}
\end{table}

\subsubsection{Quantitative result in fine segmentation stage} Table~\ref{tab1} gives a quantitative result of fine segmentation model. Note that we don't keep the largest connected component in each case.

\subsubsection{Quantitative result on independent private validation set} Table~\ref{tab2} gives a quantitative result of \textbf{AirwaySeg} on independent private validation set. Note that we keep the largest connected component in each case.

\begin{table} \centering
\caption{Quantitative results of \textbf{AirwaySeg}.}\label{tab2}
\begin{tabular}{|l|l|l|l|l|l|}
\hline
 & Dataset & Scans & Dice & False Negative & False Positive\\
\hline
AirwaySeg & online validation & 50 & 0.914±0.040 & 0.079±0.042 & 0.090±0.066\\
\hline
\end{tabular}
\end{table}

\subsubsection{Qualitative result on independent private validation set} 
\begin{figure} \centering
\includegraphics[width=\textwidth]{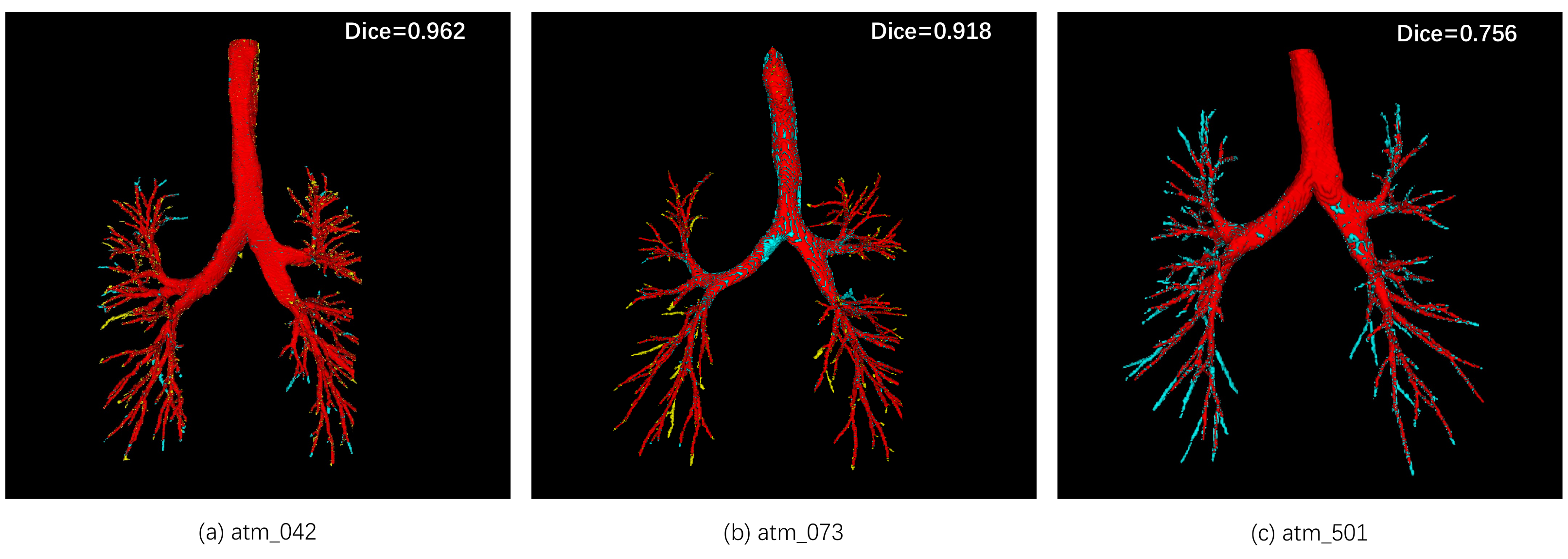}
\caption{Pulmonary airway segmentation of the proposed \textbf{AirwaySeg} method. (a) the case with dice score higher than the average dice score. (b) the case with dice score close to the average dice score. (c) the case with dice score lower than the average dice score. \textbf{Yellow} is false negative. \textbf{Cyan} is false positive.} \label{fig2}
\end{figure}

Fig.~\ref{fig2} gives a qualitative result of \textbf{AirwaySeg} on independent private validation set. We observe that Fig.~\ref{fig2} (c) only has false positive voxels. It is because nnU-Net may be over-fitting to the training set.

%
%
%
%

\end{document}